\documentclass[entropy,review,accept,moreauthors,pdftex,12pt,a4paper]{mdpi}

\lastpage{6092}
\doinum{10.3390/e17096072}
\pubvolume{17}
\pubyear{2015}
\externaleditor{Academic Editor: {Stefano Pirandola}}
\history{Received: 16 June 2015 / Accepted: 27 August 2015 / Published: 31 August 2015}

\usepackage{graphicx,bbm}
\usepackage{soul}
\usepackage{upgreek}
\usepackage{booktabs}
\usepackage{multirow}
\graphicspath{{./Figures/}}

\usepackage{enumitem}
\setitemize{parsep=0pt,itemsep=0pt,leftmargin=.4in}
\setenumerate{parsep=0pt,itemsep=0pt,leftmargin=.4in}

\Title{Distributing Secret Keys with Quantum Continuous Variables: Principle, Security and Implementations}

\Author{Eleni Diamanti $^{1,}$* and Anthony Leverrier $^{2}$}

\address{%
$^{1}$ Laboratoire Traitement et Communication de l'Information, CNRS, Telecom ParisTech, \linebreak 23 avenue d'Italie, Paris 75013, France\\
$^{2}$ Inria, EPI SECRET, B.P. 105, Le Chesnay Cedex 78153, France; {E-Mail: anthony.leverrier@inria.fr}}

\corres{E-Mail: eleni.diamanti@telecom-paristech.fr; Tel.: +33-014-581-7114.}

\abstract{The ability to distribute secret keys between two parties with information-theoretic security, that is regardless of the capacities of a malevolent eavesdropper, is one of the most celebrated results in the field of quantum information processing and communication. Indeed, quantum key distribution illustrates the power of encoding information on the quantum properties of light and has far-reaching implications in high-security applications. Today, quantum key distribution systems operate in real-world conditions and are commercially available. As with most quantum information protocols, quantum key distribution was first designed for qubits, the individual quanta of information. However, the use of quantum continuous variables for this task presents important advantages with respect to qubit-based protocols, in particular from a practical point of view, since it allows for simple implementations that require only standard telecommunication technology. In this review article, we describe the principle of continuous-variable~quantum \scalebox{1.03}{key distribution, focusing in particular on protocols based on coherent states. We discuss} the security of these protocols and report on the state-of-the-art in experimental implementations, including the issue of side-channel attacks. We conclude with promising perspectives in this research field.}



\begin{document}

\section{Introduction}

In a seminal result in 1984, Bennett and Brassard showed that it is possible for two parties to distribute a secret key in a way that is unconditionally secure against any adversary, even a quantum one \cite{BB84}. This fundamental primitive, namely quantum key distribution (QKD), is of great importance for many cryptographic tasks, such as one-time pad encrypted secure communication \cite{Sha:BellTech48} or message authentication \cite{Por:ieeeit14}. It has been thoroughly studied both in theory and in practice; indeed, the rapid progress in the field has enabled the distribution of secret keys with information-theoretic security over deployed optical fiber networks \cite{PPA:njp09,SFI:oe11}, and QKD systems are available on the market~\cite{IDQ}. The two communicating parties of a QKD protocol \cite{SBC:rmp09}, Alice and Bob, can in principle share an information-theoretic secret key after the exchange of a large number of quantum signals through a physical channel, known as a quantum channel, which is subject to eavesdropping, and additional information sent on a public, but authenticated classical channel. After Alice and Bob have agreed on a set of non-commuting quantum operators, they can encode some information into these variables: any attempt by the eavesdropper, Eve, to recover this information necessarily disturbs the transmitted quantum states and is discovered after random sampling of a fraction of Alice and Bob's correlated data.

In most commonly-used QKD systems, the key information is encoded on properties of single photons, and thus, specific components for single-photon detection are required. The quest for high-performance quantum key distribution systems in the last few years has led to several successful demonstrations based on these discrete-variable or distributed-phase reference protocols \cite{TNZ:natphoton07,RPH:njp09,DYD:apl10,KLH:natphoton15}. There~exists, however, a different type of protocol, in which information is carried by properties of light that are continuous, such as the values of the quadrature components of a coherent state. The use of such continuous-variable quantum information carriers, instead of qubits, constitutes a powerful alternative approach for QKD and more generally for quantum information processing \cite{WPG:rmp12}. From a practical point of view, for instance, continuous-variable (CV) QKD protocols present the major advantage that they only require standard telecommunication technology, and in particular, instead of dedicated photon-counting technology, they use coherent detection techniques widely used in classical optical communications. It is important to emphasize that there is a significant conceptual difference between these protocols and the standard BB84 protocol proposed by Bennett and Brassard \cite{BB84} and other discrete-variable protocols, even if the latter use coherent states: as we will see in detail in the following sections, information is encoded on non-orthogonal states, which captures the quantum nature of the CVQKD protocols; however, entirely different degrees of freedom are used in this case. This brings the need for different security proof techniques while at the same time opening the way to very practical implementations.

In the following, we begin by describing, in Section \ref{sec:principle}, the principle of CVQKD protocols focusing in particular on protocols using Gaussian modulation of coherent states. We then proceed in Section \ref{sec:security} with an overview of the current status of security proofs for such protocols. In Section \ref{sec:implementations}, we discuss the implementations of CVQKD protocols, including the first long-distance experiments of quantum key distribution using continuous variables, and in Section \ref{sec:side-channels}, we provide a brief overview of theoretical and experimental studies on the security of CVQKD systems in the presence of practical imperfections and side channels. Finally, in Section \ref{sec:conclusions}, we provide a comprehensive presentation of major challenges and perspectives in the field. Our goal in this review article is not to describe exhaustively all of available CVQKD protocols and implementations, but to focus on specific, well-understood examples to facilitate the understanding of the main ideas behind this approach for quantum key distribution.


\section{Principle of CVQKD with Coherent States}\label{sec:principle}

By definition, all CVQKD protocols encode information in the quadratures of the quantized electromagnetic field. This information is then recovered thanks to coherent detection techniques, in~particular homodyne (or heterodyne) detection of those quadratures.
From this perspective, the main distinction between discrete-variable and continuous-variable protocols lies in the detection technique that is employed: single-photon detection for the former and homodyne (or heterodyne) for the latter.

A number of CV protocols has been proposed in the literature and depend on the choice of states that are prepared: single-mode coherent or squeezed states, two-mode squeezed states; on the choice of modulation for single-mode states, Gaussian or non-Gaussian; on the choice of detection, homodyne or heterodyne; and finally, on the type of error correction (or else, {reconciliation}), direct or reverse.
Of~course, some of these protocols are easier to implement, and some have better security proofs than others. In this review, we will mainly focus on the simplest ones, which are also the best understood ones, namely one-way protocols using a Gaussian modulation. Other protocols have been investigated in the literature: two-way protocols \cite{PML08,WOP14}, protocols with a non-Gaussian modulation \cite{LKL04,HL06,LG09,SL:njp10,LG11} or post-selection \cite{SRL:prl02}; but, their security analysis is less advanced, and we will not consider them further in this short review.

As usual with QKD, a given protocol has two possible implementations, {prepare and measure} (PM) or {entanglement based} (EB), which are known to be equivalent in the case of Gaussian protocols \cite{GCW03}. In~the first case, Alice simply prepares and sends Gaussian states to Bob, who measures them with coherent detection; in the second version, Alice generates bipartite entangled states, measures the first half and sends the second half to Bob, who measures it. As long as Alice's lab and preparation is trusted, both variants have the same security. More precisely, the security of the PM version reduces to that of the EB protocol. For this reason, it is only necessary to analyze the security of EB QKD protocols.

Implementations, on the other hand, are usually simpler for PM protocols. The simplest CVQKD protocol is certainly GG02 introduced by Grosshans and Grangier in 2002 \cite{GG:prl02}, or its variant with heterodyne detection \cite{WLB04}.
We now describe the rough outline of this protocol. A much more detailed description can be found elsewhere \cite{Lev:prl15}, but is out of the scope of this paper.
The protocol consists of four main steps: (i) state distribution and measurement; (ii)~error reconciliation; (iii) parameter estimation; and (iv) privacy amplification. Note~that historically, parameter estimation used to be applied before error correction, but the novel order turns out to be more~efficient.

{(i) State distribution and measurement}:
Alice prepares a large number of coherent states $|\upalpha_1\rangle, \ldots, |\upalpha_N\rangle$, where $\upalpha_i$ are independent and identically distributed complex Gaussian variables $\mathcal{N}_{\mathbbm{C}}(0, V_0)$ with variance $V_0$. Depending on the protocol (homodyne or heterodyne), Bob measures either a random quadrature ($x$ or $p$) for each state and informs Alice of his choices or both quadratures.
Bob then obtains a list of $N$ or $2N$ real-valued numbers corresponding to his measurement outcomes. Alice has also access to her own list of data (she keeps only the relevant quadrature values if Bob performed a homodyne detection).
Denote the respective lists of Alice and Bob by $x = (x_1, \ldots, x_n)$ and $y =(y_1, \ldots, y_n)$ (where $n$ is either $N$ or $2N$).

{(ii) Error reconciliation}: The protocol achieves in general better performance with {reverse reconciliation} \cite{GVW:nat03} (except at very short distances \cite{JEK:pra14}): this means that Bob's string corresponds to the raw key, and Alice tries to guess its value. To achieve that objective, Alice and Bob use classical error correction techniques. More precisely, Alice and Bob agree on a linear error-correcting code before the protocol starts, and Bob sends to Alice the value of the syndrome of $y$ for this code. To recover $y$, Alice~simply needs to {correct} $x$, that is to decode in the coset code defined by the syndrome she received.

{(iii) Parameter estimation}: This step is useful to obtain an upper bound on the information available to Eve. For CVQKD protocols, this typically requires estimating the covariance matrix of the bipartite state shared by Alice and Bob. Once this estimate is obtained, Alice and Bob can compute the size $\ell$ of a secure key that they can extract from their state.

{(iv) Privacy amplification}: Alice and Bob apply a random universal hash function to their respective (corrected) strings and obtain two strings $S_A$ and $S_B$ of length $\ell$.

Variants of this protocol can differ in the type of states that are prepared (coherent, squeezed or even thermal) and in the detection (homodyne or heterodyne), but the main steps of the protocol remain basically identical.


\section{Security Analysis} \label{sec:security}

In this section, we address the security of CVQKD protocols with the assumption that Alice and Bob's labs, and equipment, are trusted. Note that this does not require that their source or detectors are perfect, but rather that their potential imperfections are well understood and can be modeled properly. For instance, Bob's detectors could have imperfect efficiency or add electronic noise. Such models can be easily incorporated into the security analysis.
However, we exclude side-channel attacks from the present analysis and will only discuss them in Section \ref{sec:side-channels}.

Security analysis for quantum key distribution protocols has evolved in a tremendous manner in the last decade. For a long time, the standard was to consider collective attacks in the asymptotic limit on infinitely long keys, and the goal was to compute the corresponding asymptotic collective key rate $K_{\mathrm{coll}}^{\mathrm{asympt}}$, given by \cite{DW:prsa05,KGR05}:
\begin{align}\label{eq:DW}
K_{\mathrm{coll}}^{\mathrm{asympt}} = I(A;B) - \upchi(B;E),
\end{align}
where $I(A;B)$ is the mutual information between Alice and Bob's measurements outcomes and $\upchi(B;E)$ is the Holevo information between Bob's string and Eve's quantum system. Note that $\upchi(B;E)$ should be replaced by $\upchi(A;E)$ for protocols with direct reconciliation. In a realistic setting, Alice and Bob cannot extract all of the information from their data, and it is usual to replace $I(A;B)$ by $\upbeta I(A;B)$, where the factor $\upbeta <1$ is the so-called {reconciliation efficiency}.
The Devetak--Winter formula, Equation~(\ref{eq:DW}), is~usually assumed to hold for continuous-variable protocols, and the challenge was therefore to compute $\upchi(B;E)$ or at least an upper bound for it. Indeed, the quantity $\upbeta I(A;B)$ can be directly observed in an~experiment.

The asymptotic limit assumption is very helpful for obtaining an upper bound on $\upchi(B;E)$. Indeed, one is in the situation where a given state can be observed a large number of times and can therefore be precisely estimated. In particular, one typically assumes that the covariance matrix $\Gamma_{AB}$ of the bipartite state $\rho_{AB}$ shared by Alice and Bob is known.
Then, using the optimality properties of Gaussian states~\cite{WGC:prl06}, one can show that $\upchi(B;E)$ is upper bounded by its value computed for the Gaussian state of covariance matrix $\Gamma_{AB}$ \cite{GC:prl06,NGA:prl06}:
\begin{align}
\label{eq-chi}
\upchi(B;E) \leq f(\Gamma_{AB}),
\end{align}
where $f$ is an entropic function depending on the symplectic eigenvalues of $\Gamma_{AB}$ and $\Gamma_{A|b}$, the covariance matrix of Alice's state conditioned on Bob's measurement result.
This last step completes the analysis of the security of one-way CVQKD protocols against collective attacks in the asymptotic limit.

More recently, a new paradigm for evaluating the security of QKD protocols has been put forward, notably by Renner \cite{Ren:PhD06}, following the universal composability framework of Canetti \cite{Can:focs01}.
In this paradigm, the QKD protocol is seen as a completely positive and trace-preserving map that takes as input an arbitrary bipartite state $\uprho_{A^N B^N}$ consisting of $N$ quantum systems, \textit{a priori} unknown to Alice and Bob, and returns a final state $\uprho_{S_A S_B E}$ where $S_A, S_B$ correspond to Alice and Bob's final keys and $E$ denotes Eve's quantum register. The aim of this framework is to assign a number, $\upvarepsilon$, to quantify the security of the protocol: $\upvarepsilon=0$ corresponding to perfect security. Moreover, we still want our notion of security to be composable (as was already the case with the Devetak--Winter approach), meaning that a protocol obtained by composing two subprotocols with respective security parameters $\upvarepsilon_1$ and $\upvarepsilon_2$ should have a security parameter $\upvarepsilon \leq \upvarepsilon_1 + \upvarepsilon_2$.
Such a requirement is achieved by taking $\upvarepsilon$ to be an upper bound on the {distance} between the protocol under study and an ideal protocol.
In particular, one can consider the trace distance between the output state produced by the protocol, $\uprho_{S_A S_B E}$ and the ideal state, which is $\uptau_{S S} \otimes \uprho_{E}$, where $\uptau_{SS} = \frac{1}{2^\ell} \sum_{s \in \{0,1\}^\ell} |s,s \rangle \langle s,s|$ describes a uniformly-chosen key of length $\ell$, identical for Alice and Bob, and where the tensor product indicates that Eve's system is completely uncorrelated with the final key:
\begin{align}
\label{trace-dist}
\frac{1}{2} \left\| \uprho_{S_A S_B E} - \uptau_{S S} \otimes \uprho_E \right\|_1 \leq \upvarepsilon.
\end{align}

Let us summarize the various notions of security proofs present in the literature from the strongest one to the weakest one:

\begin{enumerate}[noitemsep]
\item {Composable security against arbitrary attacks}, if one can bound the trace distance of Equation~(\ref{trace-dist}), without any restriction on the input state $\uprho_{A^N B^N}$ of the protocol.
\item {Composable security against collective attacks}, if one can bound the trace distance of Equation~(\ref{trace-dist}) under the restriction that the input state is identically and independently distributed, \emph{i.e.},~$\uprho_{A^NB^N}~=~\uprho_{AB}^{\otimes N}$.
\item {Security against collective attacks in the asymptotic limit} of infinitely many uses of the channel, if~one can compute an upper bound on the Holevo information, $\upchi(B;E)$ from Equation~(\ref{eq:DW}), between the raw key and the adversary, assuming that the quantum state shared by Alice and Bob is known. In the case of CV protocols, one only needs to assume that the covariance matrix of the state is known.
\end{enumerate}

Let us denote the respective secure key rates (final key length $\ell$ divided by the number $N$ of channel uses) for these three notions of security by $K^{\upvarepsilon}(N), K^{\upvarepsilon}_{\mathrm{coll}}(N)$ and $K^{\mathrm{asympt}}_{\mathrm{coll}}$. The first two quantities, which involve (smooth) conditional min-entropies (defined by Renner in \cite{Ren:PhD06} and later extended to infinite-dimensional quantum systems \cite{FAR11,BFS11}), include finite-size effects and, therefore, depend on $N$.
On the other hand, the asymptotic key rate is independent of $N$.
Since $K^{\upvarepsilon}(N) \leq K^{\upvarepsilon}_{\mathrm{coll}}(N) \leq K^{\mathrm{asympt}}_{\mathrm{coll}}$, the main question is to determine whether $K^{\upvarepsilon}(N)$ indeed converges to $ K^{\mathrm{asympt}}_{\mathrm{coll}}$ in the asymptotic limit and at which rate.
Often, in the literature, one can read that computing the value of $K^{\mathrm{asympt}}_{\mathrm{coll}}$ is sufficient because de Finetti-type reductions, such as \cite{RC09,LGR13}, show that the same value also holds for arbitrary attacks. The situation is unfortunately not so simple: de Finetti reductions only make sense in the composable setting, and in general, computing the Devetak--Winter formula is not sufficient to claim security against arbitrary attacks in the finite-size setting.

In general, one can either prove security against arbitrary attacks directly, for instance using the uncertainty principle as in \cite{FBB12, fur14} following the results from \cite{FBT14}, or one can first establish security against collective attacks (in the finite-size regime) and obtain a security claim against arbitrary attacks (with a worse value of $\upvarepsilon$) using a de Finetti reduction \cite{RC09,LGR13}. So far, this second approach was only applied for the protocol \cite{WLB04} with coherent states and heterodyne detection \cite{Lev:prl15}.

\begin{table}[H]
\centering
\footnotesize
\caption{Current security status of the main one-way continuous-variable quantum key distribution (CVQKD) protocols. PM, prepare and measure.}\label{tab:status}
\begin{tabular}{ccccc}
\toprule
\multirow{2}{*}{\textbf{Protocol}} & \textbf{(PM) State} & \multirow{2}{*}{\textbf{(PM) Modulation}} & \textbf{Bob's} & \multirow{2}{*}{\textbf{Best Currently-Available Security Proofs}} \\
 & \textbf{Preparation} & & \textbf{Measurement} & \\
\midrule

 \multirow{3}{*}{\cite{CLV:pra01}} & \multirow{3}{*}{squeezed} & \multirow{3}{*}{Gaussian} & \multirow{3}{*}{homodyne} & Finite-size \cite{FBB12, fur14}\\
 & & & & $K^{\upvarepsilon}(N) > 0$ for practical $N$\\

 & & & & $\lim_{N \to \infty} K^{\upvarepsilon}(N) < K^{\mathrm{asympt}}_{\mathrm{coll}}$\\

 \hline

 \multirow{3}{*}{ \cite{WLB04}} & \multirow{3}{*}{coherent} & \multirow{3}{*}{Gaussian} & \multirow{3}{*}{heterodyne} & Finite-size \cite{Lev:prl15}\\
 & & & & $K^{\upvarepsilon}_{\mathrm{coll}}(N) \approx K^{\mathrm{asympt}}_{\mathrm{coll}}$ for practical $N$\\
 & & & & $K^{\upvarepsilon}(N) =0 $ for practical $N$ \cite{LGR13}\\

\hline

\cite{GG:prl02} & coherent & Gaussian & homodyne & asymptotic collective \cite{GC:prl06,NGA:prl06,PBL08}\\

 \cite{UG15} & coherent & Gaussian 1D & homodyne & asymptotic collective \cite{UG15}\\

\cite{GC09} & squeezed & Gaussian & heterodyne & asymptotic collective \cite{PGB09}\\

 \cite{Fil08} & thermal & Gaussian & homo/heterodyne & asymptotic collective \cite{UF10,WPL10,WPR12}\\
 \hline

 \multirow{2}{*}{ \cite{MUL12}} & \multirow{2}{*}{squeezed} & Gaussian + & \multirow{2}{*}{homodyne} & \multirow{2}{*}{asymptotic collective \cite{MUL12}}\\
 && additional Gaussian &&\\

 \hline

  \multirow{2}{*}{ \cite{FC12,WRS13}} & \multirow{2}{*}{coherent} & \multirow{2}{*}{Gaussian } & homo/heterodyne + & \multirow{2}{*}{asymptotic collective \cite{FC12,WRS13}}\\
 & & & Gaussian post-selection &\\


\bottomrule
\end{tabular}


\end{table}

That being said, the behavior of the quantity $K^{\mathrm{asympt}}_{\mathrm{coll}}$ is still interesting, because it allows us to compare the various protocols and to understand the effect of losses and noise on the secret key rate. Moreover, it is reasonable to think that the proof technique of~\cite{Lev:prl15} can be generalized to most protocols with a Gaussian modulation and that the composable secret key rate $K_{\mathrm{coll}}^\upvarepsilon(N)$ valid against collective attacks will converge to the asymptotic key rate for reasonable values of $N$.
For this reason, computing an upper bound on the Holevo information $\upchi(B;E)$, that is a lower bound on $K^{\mathrm{asympt}}_{\mathrm{coll}}$, is an important first step in security proofs for continuous-variable QKD.

In Table \ref{tab:status}, we summarize the {current state-of-the-art} for security proofs for CVQKD with a Gaussian modulation. Two protocols have complete security proofs in the composable security framework. Their~entanglement-based version consists of preparing $N$ two-mode squeezed states and measuring each mode with either homodyne detection (of a randomly-chosen quadrature) for \cite{CLV:pra01} or with heterodyne detection for \cite{WLB04}.
The proof techniques are quite different: the security of \cite{CLV:pra01} is based on an entropic uncertainty principle for continuous variables \cite{FBT14}, while the security of \cite{WLB04} is obtained in two steps (security against collective attacks followed by a reduction from general attacks).
Despite this success, improvements are still called for. Indeed, in the first case of \cite{CLV:pra01}, the security proof provides a positive secret key rate $K^\upvarepsilon(N)$, which is positive for reasonable values of $N$, but unfortunately, the key rate does not converge to $K^{\mathrm{asympt}}_{\mathrm{coll}}$ for $N \to \infty$, which indicates that either the true secret key rate is overestimated by $K^{\mathrm{asympt}}_{\mathrm{coll}}$ or, more likely, that the proof technique should be improved.
On the other hand, in~the case of the heterodyne protocol with coherent states \cite{WLB04}, the quantity $K^\upvarepsilon(N)$ converges to the asymptotic value $K^{\mathrm{asympt}}_{\mathrm{coll}}$, corresponding to a Gaussian attack, but the convergence is too slow to obtain a positive key rate for a reasonable block size $N$. The main issue lies in the reduction from general to collective attack using either de Finetti's theorem \cite{RC09} or the post-selection technique \cite{LGR13} (also known as a de Finetti reduction).
We insist on the fact that these limitations might be due to insufficient proof techniques. Indeed, it is quite possible that better security proofs will be found and establish that $K^\upvarepsilon(N)$ converges to $K^{\mathrm{asympt}}_{\mathrm{coll}}$ for reasonable block lengths. We believe that this is certainly the most pressing issue in the theoretical study of CVQKD.

The security of the other protocols in the finite-size regime is less clear, and only the asymptotic key rate $K^{\mathrm{asympt}}_{\mathrm{coll}}$ is known. Applying the tools of \cite{FBB12, fur14} or \cite{Lev:prl15} is not straightforward in these cases, because squeezed states and homodyne detection seem to be required in order to use the entropic uncertainty relation, and the protocol must be sufficiently symmetric for the analysis of \cite{Lev:prl15} to go through. Establishing the composable security of these protocols remains an open question.

To conclude this section, we note that even if security proofs improve a lot in the next few years, finite-size effects will still remain an important issue for CVQKD. Indeed, the best case scenario would be that Gaussian attacks are optimal for all of these protocols, which would imply that Alice and Bob need to estimate the covariance matrix of the state they share. In particular, they would need to compute bounds on the quantum channel parameters (see Section \ref{sec:implementations} for details), a task that necessarily requires many data in the long distance (high loss) scenario.
The consequence is that very large block lengths and, therefore, extremely stable optical setups will be necessary to obtain composable security in experimental~implementations.


\section{Experimental Implementations}\label{sec:implementations}

In the previous sections, we have seen that continuous-variable QKD protocols may vary in terms of required resources, in particular for state preparation (squeezed or coherent states) and detection techniques (homodyne or heterodyne). These choices are of great importance for the security achieved by the corresponding implementations (see Table \ref{tab:status}), but also affect their performance, which is typically quantified by the maximal distance over which secret keys can be generated and the rate of their production. Another important choice is the medium used for the transmission of the quantum keys, namely optical fiber or free space, which depends on the targeted application of the QKD implementation. It is interesting to remark here on a historical note that initial proposals for CVQKD necessitated the use of squeezed states and were burdened by a 3-dB loss limit \cite{CLV:pra01,SKL:prl02}, which greatly limited their practical interest in any realistic communication scenario. Later, however, protocols using coherent states appeared (GG02) \cite{GG:prl02}, and the 3-dB limit was lifted \cite{SRL:prl02,GVW:nat03}. These results enhanced considerably the interest in the use of continuous variables for QKD and were at the basis of a series of works that led to ever better performing systems.

As in discrete-variable QKD, PM CVQKD protocols are in general easier to implement in practice. We describe in the following in some detail PM fiber optic implementations of the GG02 protocol, whose principle and security were discussed in Sections \ref{sec:principle} and \ref{sec:security}, respectively. This protocol is particularly \mbox{interesting from a practical point of view, since it merely necessitates the generation of coherent states,} their modulation in phase space and the detection of the quadratures of the received states using {homodyne (or heterodyne) techniques. The components required to achieve these functionalities are} readily available at a telecommunication wavelength, which is suitable for operation with fiber optic~ systems.

\begin{figure}[H]
\centering
 \includegraphics[width=0.75\textwidth]{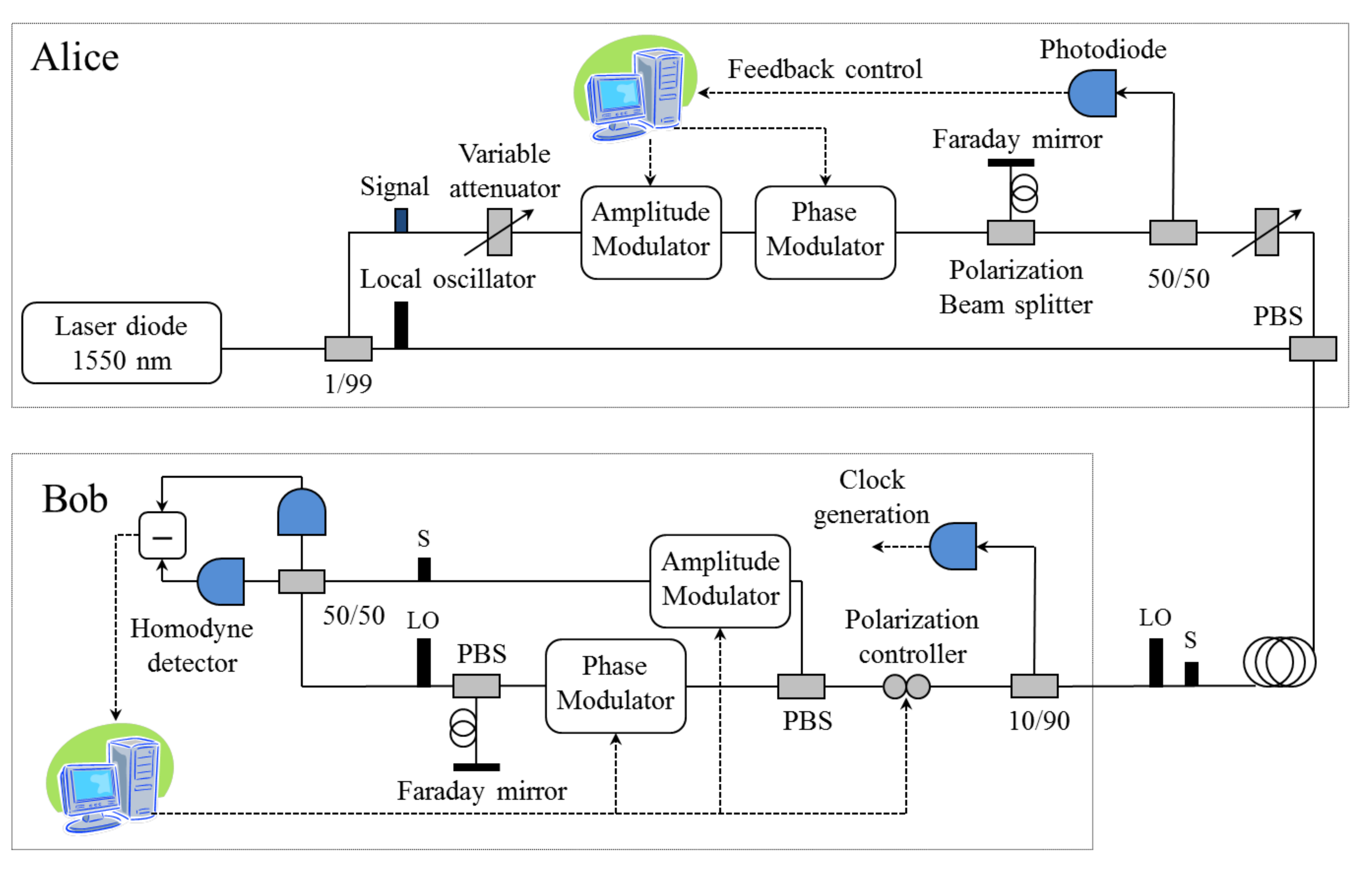}
 \vspace{6pt}
 \caption{Optical layout of a fiber optic CVQKD system implementing the GG02 \cite{GG:prl02}
 protocol with homodyne detection.}
 \label{fig:opticalsetup}
\end{figure}

The optical configuration for performing this protocol is shown in Figure~\ref{fig:opticalsetup}. In this scheme, the~signal and phase reference (or {local oscillator}) that is necessary for performing the coherent detection are generated from a laser diode source at Alice's site. The signal is modulated in amplitude and phase following a Gaussian distribution as required by the protocol and then attenuated at a suitable modulation variance level. It is also multiplexed both in time and in polarization with the local oscillator before entering the quantum channel. At Bob's site, the two signals are demultiplexed using, respectively, a delay line and a polarization beam splitter and superposed in time to interfere on a shot noise-limited balanced pulsed homodyne detector. The quadrature selection required by the GG02 protocol is performed by the phase modulator placed in the local oscillator path. The setup is completed by several active feedforward and control elements, which provide the necessary synchronization and stability conditions for performing the quantum key distribution.

The described system realizes the first part, namely {(i) state distribution and measurement}, of the full GG02 protocol described in Section \ref{sec:principle}; the remaining post-processing parts, namely {(ii) error reconciliation}, {(iii) parameter estimation} and {(iv) privacy amplification}, and, in particular, the first two, require sophisticated computational algorithms, as we will discuss further below.

The initial realization of the optical setup of Figure~\ref{fig:opticalsetup} was used in the European SECOQC QKD network \cite{PPA:njp09}, which was deployed over installed optical fibers and integrated various QKD \linebreak technologies \cite{PPA:njp09,FDD:njp09}. It was also used in a field test of a point-to-point classical symmetric encryption link with fast key renewal provided by the quantum layer, which demonstrated the reliability of the CVQKD system operation over a long period of time in a server room environment \cite{JKD:oe12}. These implementations, together with a few others \cite{QHQ:pra07,LBG:pra07,DZV:oe09,SZT:pra10}, were suited for securing communications in metropolitan area size networks (involving distances up to 25 km) with high-speed requirements. Although there are several interesting applications of short-range experiments, from a quantum information network point of view, it is important to be able to extend the communication distance beyond this limit. In discrete-variable QKD implementations, the distance limitation is essentially determined by the characteristics of single-photon detectors, and in particular their dark counts. In CVQKD, it used to be the efficiency of the complex post-processing techniques that limited the range. Although this is no longer the case, it is instructive to understand the origin of this limitation: the efficient reconciliation of correlated Gaussian variables is in fact hard, especially at low signal-to-noise (SNR) ratios, which are inherent in long-distance experiments, hence reducing the $\upbeta$ factor introduced in Section \ref{sec:security}. An an indicative value, $\upbeta = 0.9$ was achieved for an SNR = 3 (at Bob's site) in the aforementioned experiments. In recent years, a~series of works successfully addressed this issue leading to the development of highly-efficient error-correcting codes at low SNR. These combine multidimensional reconciliation techniques \cite{LAB:pra08} with efficient multi-edge low density parity check (LDPC) codes \cite{JKL:pra11} and can also be optimized for short distances \cite{JEK:pra14}. With these codes, it is possible to reach, for example, an efficiency $\upbeta = 0.96$ at SNR~=~0.075, opening the way to experiments over significantly longer distances.
We note that with the help of non-binary LDPC codes, similar efficiencies can also be obtained for higher values of the SNR~\cite{GHD:arxiv14}.

In addition to error correction, the parameter estimation procedure is also crucial for the extraction of the secret key in practice. For the optical setup of Figure~\ref{fig:opticalsetup}, the relevant experimental parameters are Alice's modulation variance $V_A$, the channel transmittance $T$ and the excess noise $\upxi$, which is the noise added by the channel beyond the fundamental shot noise and corresponds to the usual quantum bit error rate found in discrete-variable QKD implementations. Both $V_A$ and $\upxi$ are typically expressed in shot noise units. The parameter $V_A$ is adjusted in real time in order to be at all times as close as possible to the SNR corresponding to the threshold of an available error correcting code, while the parameters $T$ and $\upxi$ need to be estimated in real time by randomly revealing a fraction of the samples. Two additional experimental parameters that are used to compute an estimate of the secret information that can be extracted from the shared data are the electronic noise $v_\mathrm{el}$ and the efficiency $\upeta$ of the homodyne detection. In the so-called realistic CVQKD scenario, these are assumed to not be accessible to Eve and are measured during a secure calibration procedure that takes place before the deployment of the system. In general, however, these parameters may be available to Eve. The parameter estimation procedure allows one to compute bounds for the eavesdropper's information, taking calibrated value uncertainties into account \cite{JKDL:pra12}.

\begin{figure}[H]
\centering
 \includegraphics[width=0.68\textwidth]{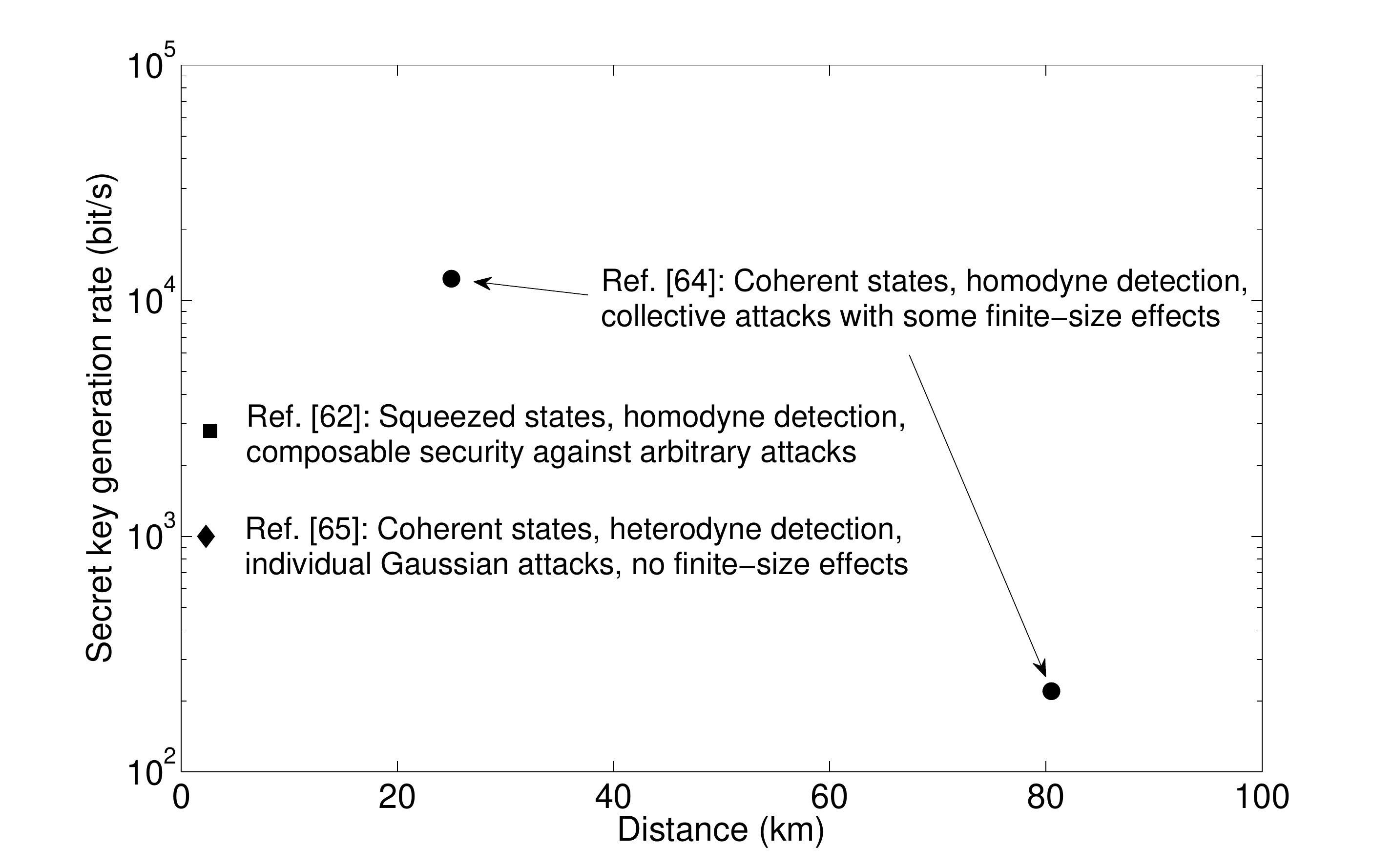}
 \vspace{6pt}

 \caption{Experimental results obtained for various CVQKD protocols offering different levels of security. The distance for the results of \cite{GHD:arxiv14} and \cite{LSS:prl05} has been calculated from data obtained with free-space experiments (expressed in dB) assuming an optical fiber with an attenuation coefficient of 0.2 dB/km, which is standard at telecommunication~wavelengths.}
 \label{fig:summary}
\end{figure}

Following error reconciliation and parameter estimation, privacy amplification allows extracting the secret information from the identical strings shared by Alice and Bob. For the scheme of Figure~\ref{fig:opticalsetup}, the~upper bound on Eve's information on the corrected string can be computed for collective attacks in both the asymptotic regime \cite{GC:prl06,NGA:prl06}, where all of the experimental parameters are assumed to be known with an infinite precision, and in the finite-size regime, where the parameters are estimated over large, but finite data pulse sets \cite{JKDL:pra12} (see Section \ref{sec:security} for rigorous security definitions). The secret key generation rates obtained with a system implementing this scheme, operating at a 1-MHz repetition rate, are shown in Figure~\ref{fig:summary} as a function of distance. Secret key generation is possible in this case at distances as long as 80 km with a data block size of $10^9$ \cite{JKL:natphoton13}. These results correspond to the current state-of-the-art in communication range for the continuous-variable QKD technology. In the same figure, we include some representative results of implementations of other CVQKD protocols. We note in particular a recent implementation involving the use of squeezed states and homodyne detectors with the goal of demonstrating composable security against arbitrary attacks (see Table \ref{tab:status}), albeit at short distances \cite{GHD:arxiv14}. Composable security against this type of attack has yet to be shown with coherent states. In this case, as was discussed in Section \ref{sec:security}, a security proof is available for heterodyne detection \cite{Lev:prl15}, but this setting was only been studied experimentally a few years ago and, therefore, does not take finite-size effects into account \cite{LSS:prl05}. Finally, we note an early implementation of the protocol employing Gaussian post-selection~\cite{SAA:pra07}, whose security proof was extended later in \cite{FC12,WRS13}.

A few remarks are in order on possible improvements for practical CVQKD implementations. First, it~is important to emphasize that thanks to the aforementioned advances, the distance limitation is no longer determined by the efficiency of the post-processing algorithms, but rather by the excess noise present in the setup and especially by the capacity to properly estimate the relevant experimental parameters, as explained above. In terms of reducing the excess noise, recent protocols based on the so-called ``noiseless amplification'' \cite{BLB:pra12,FC:pra12,WRS13} might be promising. Efficient parameter estimation over large data blocks, which requires a very stable experimental setup, plays a role not only for the distance, but also for achieving composable security, as discussed in Section \ref{sec:security}, and for increasing the secret key generation rate. Indeed, in the implementation of the GG02 protocol described above, a big fraction of the light pulses was used for this process \cite{JKL:natphoton13}; this fraction can be reduced by improving the hardware stability and, hence, by enabling the estimation of experimental parameters over larger blocks. Furthermore, the secret key generation rate may be increased by increasing the initial repetition rate of the experiments. This necessitates shortening the pulse duration and the time-multiplexing data sampling period, increasing the homodyne detection bandwidth \cite{HFW:cpl12} and performing faster error correction on multiple devices, up to the capacity of the network link used for the transmission of the classical data.


\section{Imperfections and Side Channels in Practical CVQKD}\label{sec:side-channels}

Bringing theoretical protocols to the realm of practical implementations unavoidably implies that some assumptions need to be made, such that real-life constraints can be satisfied. This may be innocuous in some cases; however, when it comes to cryptographic applications where rigorous security proofs are required, such assumptions may have dramatic consequences for the security obtained in practice. This~is of course true for quantum key distribution implementations, as well, especially as this technology is reaching a certain maturity. Let us consider, for example, the CVQKD implementation of the GG02 protocol described above. While we saw that a phase reference, the local oscillator (LO), is necessary for the implementation, in fact, this signal does not appear in any way in the description of the protocol and in the security proof. The implicit assumption made then is that Eve does not tamper with the LO; under this assumption, the security proof holds. In reality, though, there is nothing preventing Eve from manipulating this strong classical signal in order to obtain information on the transmitted key. Indeed, this is possible, as we will see below. This is a simple example of a so-called side-channel attack, which~illustrates that it is crucial to consider the practical security of CVQKD implementations.

In order to address this issue of great practical relevance, one solution is to consider exhaustively all of the possible discrepancies between the underlying theoretical model and the actual implementation, to take into account the assumptions due to experimental requirements or imperfections and to refine the model accordingly. This approach has been pursued extensively for discrete-variable QKD, where powerful side-channel attacks have been demonstrated, in particular against commercial \linebreak systems~\cite{ZFQ:pra08,XQL:njp10,LWW:natphoton10}. In~CVQKD, this process involves developing better models for the state preparation, the local oscillator manipulation and the detection stages of the implementation. We summarize below a few concrete examples of security issues that have been studied in recent years.

\subsection{State Preparation}

In practical CVQKD systems, the modulation applied to the signal according, for instance, to the GG02 protocol can only approach the Gaussian modulation required in theory. Indeed, a Gaussian distribution is not only continuous, but unbounded and, therefore, cannot be exactly achieved, since an infinite amount of randomness would be required. Using a bounded, discrete approximation, it is possible to show that the impact on security is not significant in practice \cite{JKDL:pra12}.

Additionally, similarly to the aforementioned realistic scenario by which the characteristics of the homodyne (or heterodyne) detector are assumed to be trusted (and hence, not controlled by Eve), it is possible to make the same assumption for the phase noise that is always present in the state prepared by Alice. It is then possible to show that precisely characterizing and calibrating this noise leads to an increased secret key generation rate \cite{JKDL:pra12}.

Finally, it is clear that obtaining information on the state prepared by Alice after modulation is valuable for the eavesdropper. To this end, the so-called Trojan horse attacks were studied and implemented in the discrete-variable QKD case \cite{GFK:pra06}. These attacks exploit back reflections coming from optical components, such as modulators induced by bright pulses sent by the eavesdropper, and~may open a substantial security breach. They are also effective against CVQKD systems, where Alice's modulators may be probed in this way \cite{KJS:qcrypt14}. Countermeasures against this type of attack include placing an optical isolator and a monitoring detector at the output of Alice's setup. The role of these components would then need to be explicitly included in the security proof of the implemented protocol.

\subsection{Local Oscillator Manipulation}

As was mentioned above, the presence of the intense phase reference signal, for which the no-cloning theorem does not apply, in the quantum channel is specific to standard CVQKD implementations and opens the way to potential security loopholes. Attacks based on the LO typically involve control of its intensity \cite{FGG:IQEC07,MSJ:pra13a}, and so, a monitoring detector at the entrance of Bob's site is useful in this case. The~eavesdropper can also exploit a subtle link between the local oscillator calibration procedure and the clock generation procedure employed in practical setups, such as the one illustrated in Figure~\ref{fig:opticalsetup}. In~this case, suitable manipulation of the LO leads to an overestimation of the shot noise by Alice and Bob, who then underestimate the excess noise present in the system and establish a key under conditions where no key could normally be securely generated \cite{JKD:pra13}. A suitable countermeasure for this attack consists of implementing a rigorous and robust real-time measurement of the shot noise \cite{JK:pra15}.
Another possible countermeasure against this threat is to generate the LO locally in Bob's lab, and preliminary results have recently been obtained in this direction \cite{QLP:arxiv15,SBC:arxiv15}.

\subsection{Detection}

The proposed side-channel attacks targeting the coherent detectors employed in CVQKD systems exploit either the nonlinear behavior of these detectors that can lead to their saturation \cite{QKA:spie13} or the dependence of the beam splitter included in both homodyne and heterodyne detectors on the wavelength of the incoming signal \cite{MSJ:pra13b,HKJ:pra14}. A wavelength filter is effective against the second attack, but a more general solution consists again in performing the real-time shot noise measurement analyzed in \cite{JK:pra15}. In fact, this countermeasure defeats all currently known attacks on the detection apparatus for CVQKD protocols with Gaussian modulation.

The security issues that we have discussed highlight the importance of refining security proofs of CVQKD protocols to consider practical imperfections as a means to bypass attacks based on improperly-modeled devices and procedures. Although this approach is of great practical relevance, it may be difficult in practice to identify all possible side channels present in experimental systems. A more radical approach to overcome side-channel attacks is the so-called device-independent QKD~\cite{ABG07,VV14}, where the security is guaranteed by the violation of a Bell inequality: intuitively, if Alice and Bob maximally violate the Clauser-Horne-Shimony-Holt (CHSH) inequality \cite{CHS69}, then they necessarily share a maximally-entangled state, and the eavesdropper cannot have any information about their measurement results. Unfortunately, an~implementation of device-independent QKD requires a loophole-free violation, a feat not yet achieved in the lab.
Interestingly, a much more practical variant, named {measurement device-independent} ({MDI}) {QKD}, is available and offers protection against all side-channel attacks targeting the detectors of the QKD implementation;~\cite{LCQ12} considered an MDI-QKD protocol using weak coherent pulses and decoy states, while~\cite{BP12} considered a MDI-QKD protocol in the entanglement-based representation with general finite-dimensional systems.
These results were recently extended to continuous variables in~\cite{POS:natphoton15,OSB:pra15}, which provide an unconditional security proof in the asymptotic limit (see also \cite{MSJ:pra14,LZX14} for a more restricted security analysis).

In MDI-QKD, Alice and Bob both prepare and send some states through quantum channels to a third party, Charlie, who performs an entangled measurement and announces his measurement result publicly. Conditioned on this classical information, Alice and Bob's data become correlated, and one might try to use them to extract a secure key.
This scheme can be interpreted as a time-reversal of a QKD protocol, where Charlie would send bipartite entangled states to Alice and Bob. In particular, the security of the key does not require that Charlie is trusted: if Charlie sends erroneous data, the correlations between Alice and Bob's data will not be sufficient to allow for the extraction of a key, and the protocol will simply abort. This means that a side-channel attack can only be applied against Alice and Bob's preparation procedures, which are typically easier to model properly than the detection stage.
In the continuous-variable version of MDI-QKD, Alice and Bob can, for instance, prepare coherent states with a Gaussian modulation and send them to Charlie, who mixes them on a balanced beam splitter, measures a different quadrature for both output modes and publicly announces his measurement results. Alice and Bob can then update their data using Charlie's information in order to obtain correlated continuous variables (see \cite{POS:natphoton15} for details).

The security of CV MDI-QKD can be analyzed by considering the entanglement-based version of the protocol. In that case, both Alice and Bob prepare a two-mode squeezed vacuum state, keep one half of their state and send the second half to Charlie. Once Charlie has measured his two modes and communicated his measurement result, Alice and Bob can apply suitable displacements to their respective modes. At this stage, they share a bipartite state $\uprho_{A^N B^N}$ (possibly correlated by Charlie or Eve's state), which they measure with heterodyne detection. This is similar to the CVQKD protocol with entanglement in the middle of~\cite{Wee13}.
As in Equation~(\ref{eq-chi}), the optimality of Gaussian states~\cite{WGC:prl06} guarantees that it is sufficient to know the covariance matrix of the state $\uprho_{A^NB^N}$ in order to obtain an upper bound on the Holevo information between Eve and the raw key, and in turn, a lower bound on $K_{\mathrm{coll}}^{\mathrm{asympt}}$.
Composable security against collective attacks, \emph{i.e.}, for bipartite states of the form $\uprho_{AB}^{\otimes N}$, can be established by adapting the proof of \cite{Lev:prl15}, and composable security against arbitrary attacks can finally be obtained thanks, for instance, to de Finetti reductions \cite{LGR13, RC09}.

In terms of practical implementations, MDI-QKD is very promising for discrete-variable protocols over long distances \cite{TYC14}; however, the obtained secret key generation rates remain currently relatively low. On the other hand, CV MDI-QKD implementations are limited in range, as Charlie needs to be located close to Alice's (or Bob's) lab and the channel between Bob (or Alice) and Charlie needs to feature small losses and, so, cannot exceed a few kilometers. However, the achievable rates in this case can be very high, within an order of magnitude from the known secret key capacity bounds \cite{PGB09,TGW:natcomm14}. This~configuration is therefore particularly interesting in a network setting with untrusted nodes for achieving high-speed secure communication over relatively short distances \cite{POS:natphoton15}.


\section{Conclusions and Perspectives}\label{sec:conclusions}

In the previous sections, we have provided an overview of the current achievements in the field of continuous-variable quantum key distribution, focusing in particular on the status of the security proofs for the various CVQKD protocols and the performance and limitations of practical fiber optic implementations using coherent states. These developments have undoubtedly established CVQKD as a major technology for performing secure quantum communications.

Some challenges for improving the performance of current systems, with respect in particular to the communication rate, the range of the implementations and the perspective of achieving composable security against arbitrary attacks in practice, have been discussed previously. Another major challenge for the widespread use of this technology for high-security applications involves the reduction of the size and cost of the corresponding implementations by means of photonic integration. Continuous-variable QKD is particularly well suited for integration using, for instance, silicon photonic chips, because of the standard components that it requires. Indeed, the first steps in this direction are currently being pursued~\cite{ZPH:cleo15}.

Furthermore, an important practical issue concerns the ability of QKD systems to be integrated into classical network infrastructures by means of wavelength division multiplexing techniques; here, again, CVQKD is a good candidate to achieve this goal, as has been shown recently \cite{QZQ:njp10,KQA:njp15}.

An impairment towards further development of CVQKD systems is linked to the local oscillator that needs to be sent over the quantum channel together with the signal in current standard implementations (see Figure~\ref{fig:opticalsetup}). Its presence leads to security breaches, as discussed in Section \ref{sec:side-channels}, but also is at the source of several practical problems in long-distance implementations, where, for instance, it prevents reaching a very low signal-to-noise ratio \cite{JKL:natphoton13}. This will be even more the case in future on-chip CVQKD experiments or systems adapted for free-space or satellite communications \cite{HPK:arxiv14,VBD:arxiv14}. Recent preliminary theoretical and experimental studies of a scheme that does not require the transfer of the local oscillator are promising \cite{QLP:arxiv15,SBC:arxiv15}, and further advances in this direction are likely to lead to important simplifications of practical CVQKD implementations.

These research directions, together with the possibility of using encoding on continuous variables for quantum cryptographic protocols beyond key distribution, such as bit commitment \cite{MML:pra10,MC:pra12, FSW:private}, secret sharing \cite{LW:pra13} or position-based cryptography \cite{QS:pra15}, will bring this technology a step closer to a wide range of applications within future quantum information networks.


\acknowledgments{Acknowledgments}

The authors thank Fabian Furrer, Raul Garc\'ia-Patr\'on, Phillipe Grangier, Fr\'ed\'eric Grosshans, Hoi-Kwong Lo and Stefano Pirandola for useful comments and Tobias Gehring for information on~\cite{GHD:arxiv14}.
We acknowledge financial support from the City of Paris through the project CiQWii, the French National Research Agency through the project QRYPTOS (grant number ANR-14-CE26-0011), and the Ile-de-France Region through the project QUIN (convention 13012333).





\conflictofinterests{Conflicts of Interest}

The authors declare no conflict of interest.

%

\bibliographystyle{mdpi}
\makeatletter
\renewcommand\@biblabel[1]{#1. }
\makeatother


\end{document}